\documentclass[twocolumn,american,csquotes]{article}
\usepackage[T1]{fontenc}
\UseRawInputEncoding
\usepackage{geometry}
\usepackage{units}
\usepackage{amsmath}
\usepackage{amsthm}

\makeatletter
\usepackage{abstract} 
\newcommand{\makeabstract}{\@ifundefined{abstractcontent}{}{\begin{abstract}\abstractcontent\end{abstract}}}
\newcommand{\makefrontmatter}{\if@twocolumn{\twocolumn[\maketitle\makeabstract\vskip2\baselineskip]\saythanks}\else{\maketitle\makeabstract}\fi}

\usepackage{dsfont}
\usepackage [latin1]{inputenc}

\def\one{\mathds{1}}

\makeatother

\usepackage{babel}
\usepackage[style=numeric-comp,sorting=none]{biblatex}
\begin{document}
\title{Information Field Theory\linebreak{}
Concepts, Applications, and AI-Perspective}
\author{Torsten Enßlin\\
{\small Max Planck Institute for Astrophysics, Karl-Schwarzschild-Str.\ 1,
85748 Garching, Germany}\\
{\small Deutsches Zentrum für Astrophysik, Postplatz 1, 02826 Görlitz,
Germany}}
\newcommand{\abstractcontent}{Information field theory (IFT) is the application of probabilistic
reasoning to fields. Physical fields are mathematical functions over
continuous spaces that exhibit certain properties of regularity, such
as limited variance and finite gradients. Inferring a field from an
observational dataset should exploit these regularities. However,
the finite number of constraints that the data provides is insufficient
to determine the infinite number of degrees of freedom of a field.
IFT enables us to derive optimal field inference algorithms that explicitly
exploit domain knowledge. These algorithms can be implemented via
Numerical Information Field Theory (NIFTy). In NIFTy, neural operator
forward models can be written and inverted probabilistically. NIFTy
thereby infers fields and their remaining uncertainties. This is achieved
using novel variational inference schemes that scale quasi-linearly,
even for ultra-high dimensional problems. This paper introduces the
basic concepts of IFT and NIFTy, highlights a few of their astrophysical
applications, and discusses their artificial intelligence (AI) perspective.
Finally, UBIK (the Universal Bayesian Imaging Kit), an emerging customisation
of NIFTy for a suite of astrophysical telescopes, is presented as
a central tool to the topic of the UniversAI conference.}
\makefrontmatter

\section{Introduction}

In astrophysics, one wants to deduce from observational data the state
of the physical fields that permeate the Universe, like the matter
density, the radiation field, the gravitation field, and the like.
From an artificial intelligence (AI) perspective, this is an ill posed
inverse problem, as fields have more unknown parameters -- actually
an infinite number, as at any location of continuous space fields
can and have different values -- than data constraints, as any dataset
is finite in size. The contemporary standard AI approach would be
to use a training data set, pairs of discretized field realizations
and corresponding data, in order to train and validate a neural network
to become a field reconstruction machine. This approach has a number
of problems, none severe enough to stop practitioners to go that route,
but all sufficient worrisome to to be mentioned:
\begin{itemize}
\item The ground truth in astrophysics is usually not accessible. All we
have is observational data, data products derived from these, and
our theoretical understanding of the cosmos. The observational images
we have from the universe are not a ground truth. Training networks
on those images will imprint the artifacts caused by imperfections
of the measurement and imaging processes. 
\item Synthetic skies can be obtained from numerical simulations, and mock
data can be generated with those. These, however, depend heavily on
the assumptions and approximations made and selection of included
effects in the simulations. The risk is that a neural network trained
on those generalizes properties of the simulations that do not conform
to reality. 
\item Many of the rules a network learns with large effort during training
are trivial for humans. For example that the sky emission is a strictly
positive quantity provides one bit of information per image pixel
for free, as the negative values are excluded for each pixel. Letting
a network to find this out by itself for a Mega-pixel image reconstruction
method basically wastes one million bits of freely available information.
Not using the available mathematical description of how a measurement
sensed a cosmic field might waste even more. Furthermore, it might
make it impossible to give accurately calibrated results since how
should the machine tell apart a variation in instrumental sensitivity
from a sky brightness variation?
\item Typical neural networks get trained for fixed input and output spaces.
But the variety of instruments, observational modes, and scientific
questions requires methods and results that can be flexibly reused.
The answer should be a representation of a physical field, from which
as much as possible the measurement modality is removed, so that it
can be used to answer various physical questions. 
\end{itemize}
These problems are well known in AI research, and ways to handle those
are increasingly used. One way is the usage of Bayesian reasoning,
the extension of logic to uncertainty \cite{cox1946probability},
in the way AI systems are set up. Another is to adapt appropriately
abstracted representations. For example the generalization of traditional
neural networks, which act on finite dimensional vectors, to neural
operators, which operate on fields, is definitively going into the
right direction for field inference. The reconstructed field should
become independent of the used computational grid as soon the resolution
of that is high enough and neural operators are a way to achieve this.
Furthermore, the inclusion of domain knowledge into the construction
of inference schemes is more often seen also at this conference. Physics
informed networks\textbf{ }are nice, they encourage solutions that
orient themselves along physical laws via physics-violation loss terms
in their objective functions \cites{2017arXiv171110561R}{2017arXiv171110566R}.
However, better would be solutions that can not violate these laws
at all by construction. Finally, it is more and more realized that
generative models are not only a good way to generate samples from
conditional probability distributions (all what Bayes theorem is about),
but also provide us with a language that makes the inclusion of effects
straight forward: Isn't physics all about building a forward model
for the universe?

Thus, current trends in AI research on field inference go into the
direction to separate data and field representations (by the usage
of neural operators), to use Bayesian logic, to quantify remaining
uncertainties, to impose physical laws, and to use generative models.
A mathematical framework for field inference that already does all
of this is information field theory (IFT, \cite{2009PhRvD..80j5005E,https://doi.org/10.1002/andp.201800127}).
IFT allows one to derive optimal field reconstruction algorithms. 

The algorithms derived with IFT can be implemented with the software
package \emph{Numerical Information Field Theory} (NIFTy, \cite{2013A&A...554A..26S,2024JOSS....9.6593E,2019ascl.soft03008A}).
As explained in the following, these algorithms can be understood
in terms of AI terminology as Variational Inference (VI) of latent
spaces of neural operator based generative networks with a physics-inspired
architecture. Thus, IFT can be regarded as a specific AI methodology
that is designed for knowledge guided inference of physical fields
\cite{ensslin2022information}.

These proceedings are organized as follows. First we discuss the mathematical
preliminaries of inferring a field from data, leading us to IFT. Given
this, ways to solve IFT problems on a computer using the NIFTy will
be discussed, in particular the ideas behind novel VI methods for
ultra-high dimensional problems. An outlook on the emerging \emph{Universal
Bayesian Imaging Kit} (UBIK, \cite{2024arXiv240910381E}) is given,
a unification of image and signal reconstruction for all kinds of
measurement devices. Finally, we conclude with some thoughts about
the convergence of IFT and other AI methodologies.

\section{Theory of Field Inference}

Any physical field $\varphi(x)$ depends on spatial and temporal coordinates
$x=(\vec{x},t)$ in a somehow smooth manner. Smoothness means that
field values at nearby locations are related. If they were not, very
often physical laws trigger an instantaneous dynamic that connects
or mixes the field values of nearby locations. If one knows the field
at $x$, the value at $x'$ is similar. How similar might be expressed
by the usage of a two point correlation function 
\begin{equation}
C_{\varphi}(x,x')=\left\langle \varphi(x)\,\varphi^{*}(x')\right\rangle _{(\varphi)},
\end{equation}
where we introduced averaging about a field distribution function
$\mathcal{P}(\varphi)$ and denote complex conjugation by an asterix.
If only this field correlation structure is know about the field,
the maximum entropy principle \cite{PhysRev.106.620,PhysRev.108.171}\textbf{
}encourages to assign a Gaussian probability description with this
covariance as a summary of the knowledge on the field,
\begin{eqnarray}
\mathcal{P}(\varphi|C_{\varphi}) & = & \mathcal{G}(\varphi,C_{\varphi})\ \equiv\ \mathcal{N}(\varphi|0,C_{\varphi})\\
 & := & \frac{1}{\sqrt{|2\pi C_{\varphi}|}}\exp\left[-\frac{1}{2}\varphi^{\dagger}C_{\varphi}^{-1}\varphi\right].
\end{eqnarray}
Here, the notion of a scalar product between two fields $\phi$ and
$\psi$ is introduced by defining $\phi^{\dagger}\psi=\int dx\,\phi^{*}(x)\,\psi(x)$.

Sometimes other constraints are known, for example the sky brightness
field $s(x)$ has to be strictly positive. This can be modeled via
suitable point wise transformations, e.g. by assigning
\begin{equation}
s(x)=s_{0}\,\exp(\varphi(x)),
\end{equation}
or the like. The exponentiation chosen here brings the often wanted
property that $s$ can easily vary over several orders of magnitude
by a relative moderate variation of $\varphi.$

From now on, the field of interest should be denoted by $s$, where
$s$ stands for signal, the quantity to be inferred. Be $I$ the collection
of all the background knowledge we have on $s$ (e.g. its correlations
structure, whether it is positive, etc.), then $\mathcal{P}(s|I)$
is a knowledge state over the space of field configurations. 

In order to learn more about a field its has to be probed by a measurement
device which delivers data $d$. The data are informative on the field
if its likelihood $\mathcal{P}(d|s,I)$ differs for different field
configurations, $\mathcal{P}(d|s,I)\neq\mathcal{P}(d|I)$. The data
might be considered to be split into a signal response $R(s):=\langle d\rangle_{(d|s,I)}$
and noise $n:=d-R(s)$, so that $d=R(s)+n$, but this is not essential
here. Essential is that data and its likelihood allow us to update
the prior field knowledge state $\mathcal{P}(s|I)$ to the Bayesian
posterior state (after the measurement)
\begin{equation}
\mathcal{P}(s|d,I)=\frac{\mathcal{P}(d,s|I)}{\mathcal{P}(d|I)}=\frac{\mathcal{P}(d|s,I)\,\mathcal{P}(s|I)}{\mathcal{P}(d|I)}
\end{equation}
with $\mathcal{P}(d|I)=\int\mathcal{D}s\:\mathcal{P}(d,s|I)$ the
so called evidence, which ensures proper normalization.

From this posterior moments of interests can be calculated, for example
the posterior mean field $m=\langle s\rangle_{(s|d,I)}$ or the posterior
uncertainty dispersion $D=\langle(s-m)\,(s-m)^{\dagger}\rangle_{(s|d,I)}$
and the like. 

Quantum and statistical field theory have provided a great number
of mathematical tools to calculate such conditional field expectation
values like Feynman diagrams, renormalization techniques and others.
In order to connect to these, we rephrase Bayesian probabilities in
a mathematical language that is closer to those field theories. For
this we introduce the information Hamiltonian $\mathcal{H}(\cdot|\cdot):=-\ln\mathcal{P}(\cdot|\cdot)$
in analogy to the Hamiltonian in statistical physics and the partition
functions $\mathcal{Z}=\mathcal{P}(d|I)$, which turns out to be the
evidence of Bayesian inference. 

With these Bayes theorem can we written as 
\begin{equation}
\mathcal{P}(s|d,I)=\frac{e^{-\mathcal{H}(d,s|I)}}{\mathcal{Z}},
\end{equation}
a form that now permits us to apply field theoretical methods. Interestingly,
this form brings it also closer to the AI language, as the joint information
Hamiltonian of data and signal is the sum
\begin{equation}
\mathcal{H}(d,s|I)=\mathcal{H}(d|s,I)+\mathcal{H}(s|I)\label{eq:additive}
\end{equation}
 of what would be called in AI a data fidelity term $\mathcal{H}(d|s,I)$
and a regularization term $\mathcal{H}(s|I)$. 

For example, in the above case of a log-normal signal $s=s_{0}\exp\varphi$,
measured by a device with a linear response and Gaussian noise, so
that $\mathcal{P}(d|s,I)=\mathcal{G}(d-R\,s,N)$, with $R$ the response
operator and $N$ the noise covariance, the joint information Hamiltonian
for the log-signal field $\varphi$ reads
\begin{eqnarray}
\mathcal{H}(d,\varphi|I) & = & \frac{1}{2}\left(d-R\,s_{0}e^{\varphi}\right)^{\dagger}N^{-1}\left(d-R\,s_{0}e^{\varphi}\right)+\nonumber \\
 &  & \frac{1}{2}\varphi^{\dagger}C_{\varphi}^{-1}\varphi+\mathcal{H}_{0}.
\end{eqnarray}
Here $\mathcal{H}_{0}$ collects all terms that are independent of
$\varphi$ and the exponential function is applied at each location
$x$ locally, $\left(e^{\varphi}\right)(x)=e^{\varphi(x)}$.

For fixed data, the minimum of $\mathcal{H}(d,\varphi|I)$ w.r.t.\ $s$
would correspond to the maximum a posteriori (MAP) estimator of $\varphi$.
In the specific example above this would be given by the solution
to the equation
\begin{equation}
0=\frac{\partial\mathcal{H}(d,\varphi|I)}{\partial\varphi}=-s_{0}e^{\varphi}R^{\dagger}N^{-1}\left(d-R\,s_{0}e^{\varphi}\right)+C_{\varphi}^{-1}\varphi.
\end{equation}

In general, this equation needs to be solved numerically, for example
by following the negative gradient, $-\frac{\partial\mathcal{H}(d,\varphi|I)}{\partial\varphi}$,
until a minimum is reached. Although this is well possible, a different
approach using generative models and variational inference (VI) is
recommended for such problems for three reasons. 

First, the MAP solution $\varphi_{\text{MAP}}=\text{argmax}_{\varphi}\mathcal{H}(d,\varphi|I)$
is often not the best from a minimal reconstruction error perspective,
where one wants to minimize the expectation value of the square error
$|\varphi-\psi|^{2}:=\int dx\,(\varphi(x)-\psi(x))^{2}$ of a reconstruction
$\psi$. For this, not the mode (or maximum) of the posterior is wanted,
but the mean $\overline{\varphi}=\langle\varphi\rangle_{(s|d,I)}$. 

Second, the mode $\varphi_{\text{MAP}}$ does (in general) not give
the mode of the signal field, $s_{\text{MAP}}\neq s_{0}\exp\varphi_{\text{MAP}}$,
due to the non-linearity of the relation of $s$ and $\varphi$. The
same holds for the means, $\overline{s}\neq s_{0}\exp\overline{\varphi}$. 

And third, the MAP solution does not provide uncertainty quantification,
as only a point estimate for the field is returned. 

What is desired is a set of points that are drawn from the posterior,
as from such, any moment of the fields $s$ and $\varphi$ can be
estimated. VI provides such samples to a good approximation. To understand
the usage of VI in NIFTy, we need to formulate field inference in
terms of standardized generative models.

\section{Generative Field Models}

Generative models are widely used in AI. A standardized generative
model $f_{\sigma}$ transforms some unstructured white noise vector
$\xi\hookleftarrow\mathcal{G}(\xi,\one)$ into a more structured signal
$s=f_{\sigma}(\xi)$. Here, $\sigma$ denotes the collection of parameters
of the model, for example the weights of a neural network that are
trained to create a signal distribution following that of some training
data. Generative models allow us to easily draw samples from the prior
$\mathcal{P}(s|I)\equiv\mathcal{G}(\xi,\one)\,|\partial f_{\sigma}(\xi)/\partial\xi|_{\xi=f_{\sigma}^{-1}(s)}$
via $\xi\hookleftarrow\mathcal{G}(\xi,\one)$ and $s=f_{\sigma}(\xi)$.

In IFT, such a training of generative models is often not necessary
as the parameters are specified by the prior knowledge. If, for example,
the field $\varphi$ has a Gaussian statistics with a priori a translation
invariant correlation structure $C_{\varphi}(x,x')=C(x-x')$, then
its power spectrum $P_{\varphi}(k):=\int dr\,C(r)\:e^{ikr}=:(\mathcal{F}\,C)(k)$,
with $\mathcal{F}$ the Fourier operator, allows one to write down
a generative model for $s=s_{0}\exp\varphi$, namely
\begin{equation}
s=f_{\sigma}(\xi)=\widehat{s_{0}}\exp\left(\mathcal{F}^{-1}\widehat{P_{\varphi}^{\nicefrac{1}{2}}}\xi\right).
\end{equation}
Here, $f_{\sigma}=\widehat{s_{0}}\circ\exp\circ\mathcal{F}^{-1}\circ\widehat{P_{\varphi}^{\nicefrac{1}{2}}}$
is a neural operator chain, $\sigma=(P_{\varphi},s_{0})$ is the collection
of all parameters of the operator network, and $\widehat{a}=\text{diag}(a)$
denotes a diagonal operator with diagonal $a$. The signal field $s$
can have a non-Gaussian statistic, in this specific case a log-normal
one. In case the power spectrum is not know a priori, it can be modeled
as well as a log-normal field with its own standardized variables
$\xi_{P_{\varphi}}$. This rewriting of fields (and their statistics)
in terms of a standardized generative models does not only facilitate
the generation of prior samples. It also helps to unify inference,
as every standardized model has the same prior, $\mathcal{P}(\xi|I)=\mathcal{G}(\xi,\one)$.
The corresponding joint information Hamiltonian for $\xi$,
\begin{eqnarray}
 &  & \mathcal{H}(d,\xi|I)\nonumber \\
 & = & \frac{1}{2}\left(d-R\,s_{0}e^{\mathcal{F}^{-1}\widehat{P_{\varphi}^{\nicefrac{1}{2}}}\xi}\right)^{\dagger}N^{-1}\left(d-R\,s_{0}e^{\mathcal{F}^{-1}\widehat{P_{\varphi}^{\nicefrac{1}{2}}}\xi}\right)\nonumber \\
 &  & +\frac{1}{2}\xi^{\dagger}\xi+\mathcal{H}_{0},
\end{eqnarray}
has now absorbed all model complexity into the data fidelity term.
The regularization term became a simple $\mathcal{L}^{2}$ norm on
the latent field excitation parameters $\xi$.

\section{Variational Inference}

Minimizing the standardized information Hamiltonian $\mathcal{H}(d,\xi|I)=\mathcal{H}(d|\xi,I)+\mathcal{H}(\xi|I)=\mathcal{H}(d|s=f_{\sigma}(\xi),I)+\frac{1}{2}\xi^{\dagger}\xi+\frac{1}{2}\ln|2\pi\one|$
provides the MAP solution for $\xi$, which is not necessarily the
MAP solution for $s=f_{\sigma}(\xi)$ due to non-linearities in the
model. More problematic than this is that the MAP solution is often
not a good estimator for deep and high dimensional models, like those
where a field and its power spectrum are to be inferred simultaneously
\cite{PhysRevD.83.105014}. The size of the phase space volume is
ignored by MAP, but such volume factors are essential for calculating
proper expectation values. 

Variational Inference (VI) improves on this. In VI a simpler parametric
probability distribution $\mathcal{Q}(\xi|p)$ with parameters $p$
is adapted to the posterior $\mathcal{P}(\xi|d,I)$ by minimizing
their relative entropy or Kullback-Leibler divergence 
\begin{equation}
\mathcal{D}_{\text{KL}}(\mathcal{Q}||\mathcal{P})=\int d\xi\,\mathcal{Q}(\xi|p)\,\ln\frac{\mathcal{Q}(\xi|p)}{\mathcal{P}(\xi|d,I)}
\end{equation}
 w.r.t.\ $p$. A convenient choice for $\mathcal{Q}$ is that of
a Gaussian with mean $\overline{\xi}$ and dispersion $D$, $\mathcal{Q}(\xi|\overline{\xi},D)=\mathcal{G}(\xi-\overline{\xi},D).$
Optimizing for $\overline{\xi}$ is feasible, as this is as large
as any latent space vector. However, optimizing w.r.t.\ to the entries
of $D$ is beyond current computer capabilities for most interesting
field inference problems due to the quadratic scaling of their number
with the number of signal representation pixels. Fortunately, it turns
out that a good approximation can be build for $D$ as a function
of $\overline{\xi}$ that can be stored and computed (in its action
and to a good accuracy) in quasi linear time,
\begin{equation}
D(\overline{\xi})\approx\left[\one+\mathcal{M}_{d|\overline{\xi}}\right]^{-1},
\end{equation}
where 
\begin{equation}
\mathcal{M}_{d|\xi}:=\left\langle \frac{\partial\mathcal{H}(d|\xi,I)}{\partial\xi}\,\frac{\partial\mathcal{H}(d|\xi,I)}{\partial\xi}^{\dagger}\right\rangle _{(d|\xi)}
\end{equation}
is the Fisher information matrix. Note that this matrix does not need
to be instantiated or be calculated explicitly. All what is needed
is that the data average is calculated analytically for the relevant
noise statistics and that the action of the resulting matrix applied
to a field can be computed numerically. Thanks to automatic differentiation
in NIFTy and its underlying algebra package JAX \cite{jax2018github},
this happens in the background. The integral in the Kullback-Leibler
divergence is approximated using samples drawn from $\mathcal{Q}(\xi|\overline{\xi},D(\overline{\xi}))$.

With this approximation, only the latent space posterior mean $\overline{\xi}$
needs to be optimized for. An algorithm that does this is \emph{Metric
Gaussian Variational Inference} (MGVI, \cite{2019arXiv190111033K}).
It has enabled the reconstruction of relative complex astrophysical
field models in the full electromagnetic spectrum, ranging from radio
to gamma rays \cite{junklewitz2016resolve,2024A&A...690A.387R,2015A&A...581A.126S,2023A&A...680A...2S}. 

An extension of MGVI is \emph{geometric Varitational Inference}\textbf{\emph{
}}(geoVI, \cite{2021Entrp..23..853F}). In geoVI a second latent space
is constructed via a coordinate transformation $y=g_{\overline{\xi}}(\xi)$
that follows geodesics of the metric $D_{\xi}$ in a Taylor approximation
around an expansion point $\overline{\xi}$. In these new coordinates
the posterior becomes approximatively normal, allowing to generate
samples from it, therefore from $\xi=g_{\overline{\xi}}^{-1}(y)$,
and ultimately from $s=f_{\sigma}(\xi)$. Thus, geoVI invokes a normalizing
flow. This flow needs not to be trained, it is derived from the analytical
structure of the used generative model. The resulting samples from
drawing from a Gaussian in $y$-space and converting them into signal
vectors via $s=f_{\sigma}(g_{\overline{\xi}}^{-1}(y))$ represent
the posterior distributions more accurately than MGVI. This has proven
to be in particular relevant for component separation problems, where
there are large degeneracies between components that are often only
solved by subtle forces of the prior (meaning gradients of $\mathcal{H}(s|I)$)
\cite{2024A&A...690A.314H}. 

The MGVI and geoVI methods take as an input a standardized generative
operator model, which can be written in NIFTy, and the observational
data that should be interpreted in the light of the model. They then
construct the Fisher information matrix and the field uncertainty
covariance $D$ as implicit (never instantiated) operators, optimize
for the latent space point $\overline{\xi}$ using stochastic samples
to perform the involved integrals over phase space, and return the
optimal $\overline{\xi}$ as well as a number of approximate posterior
samples $(\xi_{i})_{i=1}^{n}$ scattered around it. From those, any
desired moment $h(s)$ of the field $s$ can be calculated via a sample
average
\begin{equation}
\langle h(s)\rangle_{(s|d,I)}\approx\frac{1}{n}\sum_{i=1}^{n}h(f_{\sigma}(\xi_{i})).
\end{equation}
We remind that our field is given via the generative model $s=f_{\sigma}(\xi)$.

\section{Astrophysical Applications}

The quasi-linear scaling of MGVI and geoVI with problem size has allowed
for tackling large reconstruction problems in astrophysics. To date,
the largest one is a reconstruction of the three dimensional Galactic
dust distribution around the sun, with $6\cdot10^{8}$ voxels being
tomographically reconstructed from starlight reddening data and stellar
positions \cite{2024A&A...685A..82E}. Other areas of application
are radio \cite{junklewitz2016resolve,2024A&A...690A.387R}, X-ray
\cite{2024A&A...684A.155W,2024arXiv241014599E}, and gamma-ray \cite{2015A&A...581A.126S,2023A&A...680A...2S}
astronomy. 

The usage of IFT and NIFTy is not yet wide spread in astrophysics,
as these requires some expert knowledge in IFT, its numerics, and
the underlying mathematics. This seems to be a threshold for many
users, despite the existence of freely available educational material
\cite{ift-page}. In order to facilitate the usage of IFT based methods,
the \emph{Universal Bayesian Imaging }Kit (UBIK) software package
is under development \cite{2024arXiv240910381E}. The idea behind
UBIK is that in the IFT framework, the problem of reconstructing a
field is modular with respect to the knowledge on the field (as encoded
in the prior), the way the field is sensed (as encoded in the likelihood),
and how the inference is performed (whether via MAP, MGVI, or geoVI).
Any combination of these elements gives rise to an algorithm tailored
to a specific measurement situation and scientific question. Furthermore,
thanks to the additivity of information, see Eq.\ \ref{eq:additive},
the fusion of data sets from different instruments is well possible.
For example if data sets $d_{1}$ and $d_{2}$ are independent measurements
of the same signal, $\mathcal{P}(d_{1},d_{2}|s,I)=\mathcal{P}(d_{1}|s,I)\,\mathcal{P}(d_{2}|s,I)$,
the corresponding joint information Hamiltonian of data and signal,
\begin{equation}
\mathcal{H}(d,s|I)=\mathcal{H}(d_{1}|s,I)+\mathcal{H}(d_{2}|s,I)+\mathcal{H}(s|I),
\end{equation}
just contains the sum of the corresponding data fidelity terms. Provided
with a suite of instrument descriptions, UBIK should enable IFT based
reconstructions for a large number of astrophysical instruments and
multi-messenger physics with those. In its current version, it is
able to handle data from the \emph{Chandra}, the \emph{eROSITA}, and
from the \emph{JWST} satellite. The inclusion of a larger suite of
radio astronomical instrument descriptions is foreseen next.

\section{Conclusion}

Information field theory (IFT, \cite{2009PhRvD..80j5005E,https://doi.org/10.1002/andp.201800127})
emerged from the need to have a mathematical theory of field inference
that works in practice, for example on astrophysical data sets. On
a first glance, IFT seems to be unrelated to most artificial intelligence
(AI) developments, except maybe Gaussian process regression. However,
the computational tools developed for IFT problems benefited from
and extended methods that were designed for AI. Thus methods used
in IFT applications converge towards those used in AI. 

Current trends in AI towards more detector independent representation
of information, e.g. via neural operators, towards a stricter adherence
to probabilistic logic and physical laws (via physics informed networks),
indicate that there is a convergent movement also on the AI side towards
IFT principles. From these observations, it is apparent that IFT is
an AI methodology that puts more emphasis on the known structure of
the physical world than pure data driven AI methods \cite{ensslin2022information}.
But the degree of data versus knowledge driven can be freely chosen,
and the inclusion of purely data driven networks into the IFT methodology
has already been demonstrated successfully \cite{2021Entrp..23..693K}.

An area where IFT might help the understanding and development of
AI methods is that the IFT concept of probability distributions over
fields is a useful language to phrase the AI problem of how to train
a network. The versatility of neural networks stems from their \emph{Universal
Approximation Theorem} \cite{hornik1989multilayer}. Any reasonable
function can be approximated by a network of sufficient width (or
depth). This theorem is, however, also a curse: Given a finite set
data, there is also an infinite number of functions that will pass
through it perfectly. How and whether the right one is chosen during
neural network training, and what this implies in terms of implied
data generalization rules, is unclear. IFT, the information theory
for fields, might provide a suitable language to address such questions.

\subsection*{Acknowledgments}

TE thanks the organizers of the UniversAI IAU conference in Athen
2025 for the kind invitation and Anton Nöbauer, Julian Rüstig, and
Lena Tauber for comments on the manuscript. Funded by the European
Union (ERC, mw-atlas, 101166905). Views and opinions expressed are
however those of the author only and do not necessarily reflect those
of the European Union or the European Research Council Executive Agency.
Neither the European Union nor the granting authority can be held
responsible for them.

\printbibliography

\end{document}